\documentclass[onecolumn]{emulateapj}

\usepackage{multirow,graphicx}

\renewcommand{\vec}[1]{\mathbf{#1}}
\newcommand{\el}{$2003~{\rm EL_{61}}$}

\begin{document}

\title{L\'{e}vy Flights of Binary Orbits due to Impulsive Encounters}
\author{Benjamin F. Collins\altaffilmark{1} and Re'em Sari\altaffilmark{1,2}}
\altaffiltext{1}{California Institute of Technology, MC 130-33,
        Pasadena, CA 91125}
\altaffiltext{2}{Racah Institute of Physics, Hebrew University, Jerusalem 91904,
Israel}
\email{bfc@tapir.caltech.edu}

\begin{abstract}
We examine the evolution of an almost circular Keplerian orbit
interacting with unbound perturbers.  We calculate the change in
eccentricity and angular momentum that results from a single
encounter, assuming the timescale for the interaction is shorter than
the orbital period.  The orbital perturbations are incorporated into a
Boltzmann equation that allows for eccentricity dissipation.  We
present an analytic solution to the Boltzmann equation that describes
the distribution of orbital eccentricity and relative inclination as a
function of time.  The eccentricity and inclination of the binary do
not evolve according to a normal random walk but perform a L\'{e}vy
flight.  The slope of the mass spectrum of perturbers dictates whether
close gravitational scatterings are more important than distant tidal
ones.  When close scatterings are important, the mass spectrum sets
the slope of the eccentricity and inclination distribution functions.
We use this general framework to understand the eccentricities of
several Kuiper belt systems: Pluto, \el, and Eris.  We use the model
of \citet{TBGE07} to separate the non-Keplerian components of the
orbits of Pluto's outer moons Nix and Hydra from the motion excited by
interactions with other Kuiper belt objects.  Our distribution is
consistent with the observations of Nix, Hydra, and the satellites of
\el and Eris.  We address applications of this work to objects outside
of the solar system, such as extrasolar planets around their stars and
millisecond pulsars.
\end{abstract}


\section{Introduction}

Several binary Kuiper belt objects (KBOs) have well-measured small
orbital eccentricities \citep{NGC07}.  \citet{SBL03} investigate
numerically the forcing of the eccentricity of the Pluto-Charon orbit
by interloping KBOs.  They find that the system almost never possesses
an eccentricity as high as the observed value of 0.003 \citep{TBGE07};
depending on the model of tidal damping used, they find median values
of $10^{-5} - 10^{-4}$.  Our goal is to develop an analytic theory that
describes the effects of a population of unbound perturbers on a
binary orbit and can be applied simply to any binary, in the Kuiper
belt or elsewhere.

The interaction of a binary system with its environment has been
studied extensively in the literature \citep{H75,SS95,Y02,MME07,SHM07}.  
One interesting context is
white dwarf-pulsar binaries, which are expected to be circular.  For
these objects pulse timing produces very accurate measurements of
their orbital motion; such measurements reveal that their
eccentricities are typically very small but finite, around
$10^{-4}-10^{-5}$ \citep{S04}.  \citet{P92} investigated the effects
of passing stars on the orbit of such a binary and found that for
Galactic pulsars, the perturbations are sub-dominant compared to the
effects of atmospheric fluctuations in the companion star.  The higher
density environment of a globular cluster however can induce an order
of magnitude higher eccentricity.  \citet{RH95} and \citet{HR96}
present a detailed account of the changes in orbital parameters for
binaries in a stellar cluster.  The work of these authors focuses on
the regime where a perturbing body interacts with the binary on
timescales longer than the orbital period of the binary.  In the
Kuiper belt, a single interaction between a binary and an unbound
object occurs over a shorter timescale than the orbital period of the
binary.  We focus on this regime, where the perturbations 
to the orbital dynamics can be approximated as discrete impulses.

The main result of this work is that we have identified the
perturbative evolution of the eccentricity and relative inclination of
a nearly circular binary orbit as a L\'{e}vy flight, a specific type of 
random walk through phase space \citep{SZF95}.  The entire
distribution function of the eccentricity and inclination is then
determined by calculating the frequency of perturbations as a function
of their magnitude.    We find a simple analytic expression for this
distribution function.

We take the following steps to arrive at 
our conclusion.  In section \ref{secSingleEnc} we
calculate the effect of one perturber on a two-body orbit, examining
separately the tidal effects of distant scatterings, close encounters
with a single binary member, and direct collisions.  We describe the
effects of many such encounters in section \ref{secBE}, and write a
Boltzmann equation that describes the distribution function of the
orbital eccentricity and the inclination of the binary relative to its
initial plane.  The quantitative description of the binary's evolution 
given by this distribution function reveals its nature as a L\'{e}vy flight.
In section \ref{secMassSpec}, we allow for a
distribution of perturbing masses and discuss the different L\'{e}vy
distributions that result.

We then use the analytic theory to examine the orbits of binary KBOs
being perturbed by the other members of the Kuiper belt.  Section
\ref{secKBB} applies our analysis to several specific Kuiper belt
binaries.  We briefly discuss the relevance of this theory to other
astrophysical systems in section \ref{secApp}, and summarize our
conclusions in section \ref{secConclusions}.

\section{A Single Encounter}
\label{secSingleEnc}

We use the following terminology to describe the geometry of the
encounter between a single perturber and a two-body orbit.  We refer
to the two bound bodies as ``the binary.''  The members of the
binary have masses $m_1$ and $m_2$, with a total mass labeled
$m_b=m_1+m_2$ and $m_1 \geq m_2$.  The position of body 2 relative to body
1 is given by $\vec r_b$, and the relative velocity by $\vec v_b$.  We
distinguish between the magnitude and direction of a vector with the
notation $\vec r_b = r_b \hat r_b$.  We assume
$v_b \approx \Omega r_b$, where $\Omega$ is the orbital frequency of
the binary.  We write the orbital period as $T_{\rm orb}=2 \pi/\Omega$.

\begin{figure}[t!]
\center
\includegraphics[angle=-90]{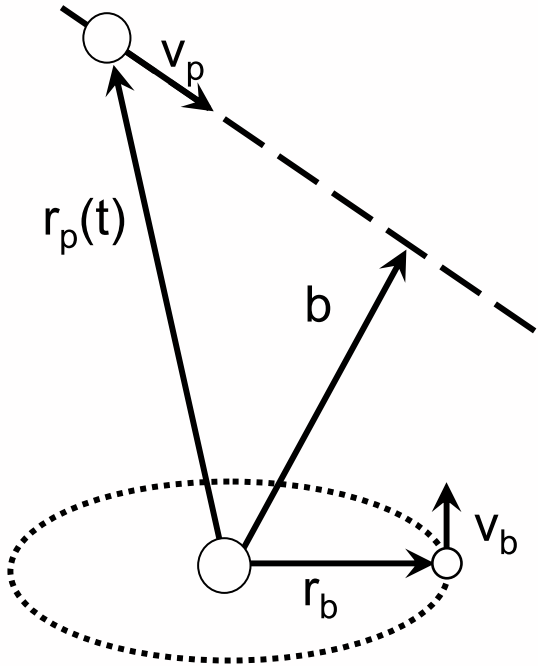}
\caption{An illustration of the notation we use to denote the geometry of each
perturbation.  The dotted line is the almost circular orbit of the binary viewed 
at an angle.  The dashed line is the path of the perturber, given by 
$\vec r_p(t)= \vec b + \vec v_p t$.}
\label{figCartoon}
\end{figure}

We label the mass of the perturber $m_p$.  The position of the
perturber as a function of time, $\vec r_p(t)$, is described by two
vectors: $\vec r_p(t) = \vec b + \vec v_p t$.  The vector $\vec b$
specifies the closest point of the perturber's trajectory to body 1,
and $\vec v_p$ is velocity of the perturber relative to body 1.  Each
encounter geometry is uniquely specified by $\vec b$ and $\vec v_p$
under the constraint $\vec b \cdot \vec v_p =0$.  Figure
\ref{figCartoon} depicts the arrangement of the vectors $\vec r_b,
\vec v_b, \vec r_p(t), \vec b,$ and $\vec v_p$.  We assume $T_{\rm
  orb} \gg b/v_p$ so that we may ignore the motion of the binary
during the interaction.  We further assume that the effects of the
gravity of the binary on the perturber are small; the perturber then
travels along a straight path with a constant $\vec v_p$.  This
assumption requires the criterion of $v_p^2 \gg G (m_b+m_p)/b$.  If
$b$ is small, the perturber may collide with a member of the binary.
In this case the assumption that the path of the perturber is
unaffected by the gravity of the binary is true under the condition
that $v_p$ is much greater than the escape velocity of that member of
the binary.  The escape velocity from body 1 is defined $v^2_{{\rm
    esc,1}}=2 G m_1/ R_1$, where $R_1$ is the radius of body 1.

We are assuming that the timescale of the interaction is much shorter than
the orbital timescale, such that the perturbation instantaneously changes the
velocities of the binary objects.
The impulse provided to a specific member of the binary is found by
integrating the acceleration caused by the perturber over its path:

\begin{equation}
\label{eqDeltaV1}
\Delta \vec v_j = \int_{-\infty}^{\infty} 
\frac{G m_p(\vec b_j + \vec v_p t)}{|\vec b_j+ \vec v_p t|^3}
 dt = 2 \frac{G m_p}{v_p}
\frac{\hat b_j}{b_j},
\end{equation}

\noindent
where the index $j$ specifies whether the impulse $\Delta \vec v_j$
and impact parameter $\vec b_j$ are 
with respect to either the primary ($j=1$) or the secondary
($j=2$).  For the primary, $\vec b_1 = \vec b$ as we have defined it
above.  For encounters with the secondary, $\vec b_2$ is related to
$\vec b$ by enforcing that it is also perpendicular to $\vec v_p$.
Thus we find $\vec b_2 = \vec b - \vec r_b + \hat v_p ( \vec r_b \cdot
\hat v_p)$.

We
consider the effects of such impulses on the full Laplace-Runge-Lenz
vector, $\vec e = (\vec v_b \times \vec H)/G m_b - \hat r$, where
$\vec H=\vec r_b \times \vec v_b$, the angular momentum per unit mass
of the binary.  The vector $\vec e$ has a magnitude equal to the
eccentricity of the orbit, and points from body 1 towards
the periapse.  It responds to a small impulse $\Delta \vec v$ according to
the formula

\begin{equation}
\label{eqEdot}
\Delta \vec e = \frac{1}{G m_b} \left[ 2 \vec r_b (\Delta \vec v \cdot \vec v_b)
- \vec v_b (\Delta \vec v \cdot \vec r_b) 
- \Delta \vec v (\vec r_b \cdot \vec v_b)
\right],
\end{equation}

\noindent
keeping terms up to linear order in $\Delta \vec v$.
Since we have assumed the binary has very small eccentricity, the
third term in equation \ref{eqEdot} is negligible compared to the
other two.  

The orbital plane of the binary is defined by the angular momentum
vector $\vec H$, and evolves according to $\Delta \vec H = \vec
r_b \times \Delta \vec v$.  The impulses affect the
direction of the angular momentum vector, and therefore alter the
orientation of the orbital plane of the binary.  We use the
two-dimensional vector $\vec i$ to denote the components of $\hat H$
in the plane defined by the initial angular momentum.  This vector,
$\vec i$, has a magnitude equal to $\sin i$, the sine of the
inclination of the binary with respect to the initial orbital plane,
and points from body 1 towards the longitude of the ascending node.

The change in relative velocity given by a general gravitational scattering 
is given by $\Delta \vec v = \Delta \vec v_2 - \Delta \vec v_1$.  The 
resulting change in the eccentricity vector is

\begin{equation}
\label{eqDeltaegeneral}
\Delta \vec e = 2 \frac{m_p}{m_b} \frac{v_b}{v_p} 
\left[ 2 \hat r_b \left( \frac{\hat b_2 \cdot \hat v_b}{b_2/r_b} 
- \frac{\hat b \cdot \hat v_b}{b/r_b} \right)
- \hat v_b \left( \frac{\hat b_2 \cdot \hat r_b}{b_2/r_b} -
\frac{\hat b \cdot \hat r_b}{b/r_b} \right) \right].
\end{equation}

\noindent 
The change in $\vec i$ is

\begin{equation}
\label{eqDeltaigeneral}
\Delta \vec i = - 2 \frac{m_p}{m_b} \frac{v_b}{v_p}
\left[ \hat v_b \left(\frac{\hat b_2 \cdot \hat n}{b_2/r_b} 
- \frac{\hat b \cdot \hat n}{b/r_b} \right) \right],
\end{equation}

\noindent
where $\hat n$ is the unit normal vector to the binary's orbital 
plane.
For both the farthest perturbers and the closest, the dependence of 
equations \ref{eqDeltaegeneral} and \ref{eqDeltaigeneral} on the impact parameter
can be simplified.  We discuss these limits in the following sections.

\subsection{Close Encounters}
\label{secVClose}

Interactions with impact parameters greater than the radius of the
primary or secondary but much less than the semi-major axis of the
binary belong to what we call the ``close-encounter regime.''  By
definition the encounters in this regime of impact parameter are much
closer to one member of the binary than the other.  As a result the
relative impulse experienced is dominated by the single impulse
delivered to that body, $|\Delta \vec v| \approx |\Delta \vec v_j|$.
The changes in $\vec e$ and $\vec i$ are then given not by the
difference of the impulses on each body, as in equations
\ref{eqDeltaegeneral} and \ref{eqDeltaigeneral}, but by the effects of
only the largest impulse.  For the change in eccentricity we find,

\begin{equation}
\label{eqEClose}
\Delta \vec e = 2 \frac{m_p}{m_b} \frac{v_b}{v_p} \frac{r_b}{b}
\left[ 2 \hat r_b (\hat b_j \cdot \hat v_b) - \hat v_b (\hat b_j \cdot \hat r_b)
\right],
\end{equation}

\noindent
and for the inclination,

\begin{equation}
\label{eqIClose}
\Delta \vec i = - 2 \frac{m_p}{m_b} \frac{v_b}{v_p} \frac{r_b}{b}
\left[ \hat v_b (\hat b_j \cdot \hat n) \right].
\end{equation}

\subsection{Distant Encounters}

For interactions where $b \gg r_b$, the impulse delivered to each
member of the binary is almost the same.  In this limit only the tidal
difference in impulse affects the eccentricity of the binary.  The
perturbation delivered to the lowest order in $r_b/b$ is

\begin{equation}
\label{eqDeltae}
\Delta \vec e = 2 \frac{m_p}{m_b} \frac{v_b}{v_p} \left(\frac{r_b}{b}\right)^2
\left[ \hat r_b \left( 4 (\hat r_b \cdot \hat b)( \hat v_b \cdot \hat b)
+ 2 (\hat r_b \cdot \hat v_p)( \hat v_b \cdot \hat v_p) \right)
  + \hat v_b \left(1 - 2 (\hat r_b \cdot \hat b)^2 - (\hat r_b \cdot \hat v_p)^2 
\right) \right].
\end{equation}

\noindent
\citet{P92} derives the special case of
this formula for interactions that take place entirely in the plane of
the binary.  This formula is also equivalent to equation A24 of
\citet{HR96}.

The change in $\vec i$ due to distant encounters is given by:

\begin{equation}
\label{eqDeltah}
\Delta \vec i = \frac{m_p}{m_b} \frac{v_b}{v_p} \left(\frac{r_b}{b}\right)^2
\hat v_b \left[ 4(\hat r_b \cdot \hat b)(\hat b \cdot \hat n)
+ 2 (\hat r_b \cdot \hat v_p)(\hat v_p \cdot \hat n) \right].
\end{equation}

\subsection{Collisions}
\label{secCollisions}

Physical collisions between perturbers and body 1 or 2 cause the orbit
to evolve impulsively.  We define collisions to be any encounters
where the impact parameter is smaller than the radius of the primary
or secondary: $b<r_1$ or $b_2<r_2$.  In this case the impulse is given
by the conservation of linear momentum of the encounter: $\Delta \vec
v = \chi (m_p/m_j) \vec v_p$, where $m_j$ is the mass of the binary
member involved in the collision ($j=1$ or $2$).  The coefficient
$\chi$ accounts for the final momentum of the perturber.  For an
inelastic collision with $m_p \ll m_j$, $\chi=1$.  If the perturber is
perfectly reflected, $\chi=2$.  The momentum loss from an impact
crater can enhance this factor above 2 depending on the properties of
the colliding bodies \citep{MNZ94}.  For simplicity we
assume that the mass of each binary
member remains unchanged after each collision.

The collisional impulse changes the eccentricity according to equation
\ref{eqEdot} and the orbital plane according to the change in angular
momentum $\vec r_b \times \Delta \vec v$. 

\section{Boltzmann Equation}
\label{secBE}

The evolution of the eccentricity and inclination (relative to the
initial orbital plane) is given by the sum of the perturbations the
binary receives as it travels through a swarm of perturbers.  From the
average properties of the perturbing population, we can calculate a
distribution function that describes the evolution of the orbit in a
statistical sense.

\subsection{Eccentricity}
\label{secBEecc}

Since the perturbation in eccentricity is a two-dimensional vector,
each component is added to the components of the existing eccentricity
vector separately.  As the binary experiences many perturbations, its
eccentricity vector travels throughout this two-dimensional space.  We
write a distribution function $f(\vec e,t)$ that describes the
probability that the binary will have an eccentricity 
in a small region $d^2 \vec e$.  Assuming isotropic perturbations, there
is no preferred longitude of periapse for the binary.  It follows that
$f(\vec e,t)=f(e,t)$ and the likelihood of finding the
eccentricity in a small range $de$ around $e$ is $2 \pi e f(e, t) de$.

We define ${\cal R}(e')$ to be the frequency at which the binary
experiences perturbations of magnitudes between $e'$ and $e'+de'$.
The frequency of perturbations with magnitudes on the order of 
$|\Delta \vec e|=e'$ is given
schematically by $e' {\cal R}(e') \sim n v b^2$, where
$n$ is the number density of the perturbers, $v$ is the speed at
which the binary encounters those perturbers, and $b$ is the
distance at which the binary encounters perturbers that cause a
perturbation of strength $e'$.  We make this calculation precise with 
the following integral:

\begin{equation}
\label{eqRofe}
{\cal R}(e')= \int \delta(| \Delta \vec e(\vec v_p,\vec b,m_p)| - e')
{\cal F}(\vec v_p,m_p) v_p \delta(\vec b \cdot \hat v_p) d^3 \vec b d^3 \vec v_p dm_p,
\end{equation}

\noindent 
where ${\cal F}(\vec v_p,m_p)$ is the phase space density per unit
mass of the perturbers.  The integral of ${\cal F}(\vec v_p,m_p)$ over
$d^3 \vec v_p dm_p$ is the number density of the perturbers.  We
assume this density is uniform in the spatial dimensions and isotropic
in velocity.  It is normalized such that the total mass density of
perturbers is given by $\rho= \int m_p {\cal F}(\vec v_p,m_p) d^3 \vec
v_p dm_p$.  The factor of $v_p$ in the integrand of equation
\ref{eqRofe} represents the velocity at which the binary encounters
perturbers.  The second delta function in equation \ref{eqRofe}
converts the volume element $d^3 \vec b$ to an element of
cross-sectional area.  The first delta function, $\delta(| \Delta \vec
e(\vec v_p,\vec b,m_p)| - e')$, restricts the integral to include only
the combinations of $\vec b$, $\vec v_p$, and $m_p$ that cause a
$|\Delta \vec e|=e'$.

The evolution of the distribution function as a result of these perturbations 
is given by a Boltzmann equation that links the rate of change of $f(e,t)$
to the interaction frequency.  We write this equation as:

\begin{equation}
\label{eqFofe}
\frac{\partial f(e,t)}{\partial t} 
= \int p(\vec e') \left[ f(|\vec e'+\vec e|) - f(e)\right] d^2\vec e'
\end{equation}

\noindent
The function $p(\vec e')$ describes the frequency per unit of 
eccentricity space ($d^2 \vec e'$) at which a binary with
eccentricity $\vec e$ is perturbed to the value $\vec e+\vec e'$.
Since there is no preferred direction for the encounters, this
function is axisymmetric, $p(\vec e') = p(e')$.  It is related to
${\cal R}(e')$ by integrating over the angular 
direction of the phase space, ${\cal R}(e') = \int p(e') e' d \omega = 2 \pi e' p(e')$.

We first derive $p(e')$ for a simple scenario: a population of
perturbers each with mass $m_p$ and velocity $v_p$.  To clarify this
derivation, we present a qualitative treatment.  The eccentricity
excited by such a perturber with an impact parameter of order $b
\gg r_b$ is about $e' \sim (m_p/m_b)(v_b/v_p)(r_b/b)^2$ (Section
\ref{secSingleEnc}).  Since the frequency of encounters with impact
parameters $b$ is proportional to $b^2$, and the size of the
perturbation $e' \propto b^{-2}$, the frequency at which the binary is
perturbed by an amount of order $e'$ is therefore a power law: 
$e'^2 p(e') \propto e'^{-1}$.   
This power law is valid from very low
$e'$, caused by the farthest possible impulsive encounter, to $e' \sim
(m_p/m_b)(v_b/v_p)$, the rare encounters with $b \sim r_b$.  We take
into account the very rare occurrence of a physical collision, which
excite eccentricities of order $e' \sim (m_p/m_j)(v_p/v_b)$, in section
\ref{secCollisions}.

Evaluating equation \ref{eqRofe} using $\Delta \vec e (\vec v_p,\vec b,m_p)$
given by equation \ref{eqDeltae} provides the exact form of 
$p(e')$ for this scenario.
We find:

\begin{equation}
\label{eqPofE}
p(e') = 
\frac{\langle C_e \rangle}{4 \pi} G \rho T_{\rm orb} \frac{1}{e'^3},
\end{equation}

\noindent 
where $T_{\rm orb}$ is the orbital period of the binary, and $\langle
C_e \rangle = 1.89$ is the average value of the angular terms of
equation \ref{eqDeltae} (see Appendix).   
We note that the frequency of perturbations depends not on
$m_p$, but only on the total mass density of perturbers.  It is also
independent of $v_p$, as the lowered effectiveness of the faster
perturbations is directly canceled by their higher frequency.
These properties are typical of distant encounters with 
binaries, as evident in earlier work on binary dynamics \citep{BHT85}.

We can generalize equation \ref{eqFofe} by including a term to
account for dissipation of the binary's eccentricity: $\partial f(\vec
e,t)/\partial t = {\rm -div}(f(\vec e,t) {\dot \vec e})$.  We restrict
our attention to mechanisms that reduce $\vec e$ at a timescale that
is independent of $\vec e$, ${\dot \vec e} = - \vec e / \tau_d$.  The
tidal dissipation of eccentricity obeys this form and is our main
motivation for including such terms.

Since $p(e')$ is a power-law, we can look for self-similar solutions
to the time-dependent integro-differential Boltzmann equation,
equation \ref{eqFofe}.  The frequency of perturbations $p(e')$ does
not depend on any special eccentricity, so the distribution function
should depend only on the time $t$.  We separate the
distribution function into three parts: the time-dependent
normalization, $F(t)$, the time-independent shape of the function,
$g(x)$, and the time-dependent eccentricity scale, $e_c(t)$.  These
quantities obey the relation $f(e,t)=F(t)g(e/e_c(t))$.  We choose the
normalization of $g(x)$ such that $\int g(x) d^2 x=1$.  We further
choose that $f(e,t)$ be normalized to 1 for all times; this constrains
the normalization function to be $F(t)=1/e_c(t)^2$.

Substituting $f(e,t)=e_c(t)^{-2}g(e/e_c(t))$ into equation \ref{eqFofe},
we find two equations.  The first specifies the time-independent
shape of the distribution as a function of the dimensionless 
parameter $x \equiv e/e_c(t)$:

\begin{equation}
\label{eqFofg}
2 g(x)
+x \frac{d g(x)}{d x}
+ \frac{1}{2 \pi} \int\int 
\frac{g(x_n)-g(x)}{|\vec x_n - \vec x|^3} d^2\vec {x}_n = 0,
\end{equation} 

\noindent 
The solution to this equation has been presented in several earlier
works, where we investigate the eccentricity distribution of the
oligarchs in a protoplanetary disk \citep{CS06,CSS07}:

\begin{equation}
\label{eqGofx}
g(x)=\frac{1}{2\pi} (1+ x^2)^{-3/2}.
\end{equation}

\noindent
This function is the two-dimensional Cauchy distribution.  
The median and mode of this distribution are
$x_{\rm med}=\sqrt{3}$ and $x_{\rm mode}=1/\sqrt{2}$.  The mean of
this distribution is formally divergent; assuming there is a maximum 
value of $x$, $x_u \gg 1$, then $x_{\rm mean}\approx 2.3 \log_{10} (0.74 x_u)$.

The eccentricity scale $e_c(t)$ is set by an ordinary differential 
equation,

\begin{equation}
\label{eqDiffeoft}
\dot e_c(t) = - e_c(t)/\tau_d + \langle C_e \rangle G \rho T_{\rm orb}/2 
\end{equation}

\noindent
We note that $\tau_d$ and the terms on the right
hand side of equation \ref{eqDiffeoft} do not need to be constant in time;
evolution of the binary ($T_{\rm orb}(t)$), the perturbing swarm
($\rho(t)$), or the damping mechanism ($\tau_d(t)$)
can be treated by including the time-dependence of these quantities.

We offer a reminder that $e_c(t)$ is the characteristic value of the
entire distribution of eccentricity that the binary may attain.  The
probability is highest that the binary will have an eccentricity near
the mode of the distribution, which is smaller than $e_c(t)$ by a
factor of 0.7.  The distribution is somewhat wide, and the confidence
levels around the median value are large.  The 66 percent confidence
interval of $x$ is $0.67-5.8$, and the 95 percent interval is
$0.23-40.0$.

Equations \ref{eqGofx} and \ref{eqDiffeoft} present a new picture of
the stochastic evolution of the binary's eccentricity.  Often the
evolution of a random variable is characterized by Brownian motion,
in which the distribution of the random variable is
set by the long term accumulation of many small perturbations.  The
typical value of such a variable grows as the square-root of time
(written $\sqrt{\langle x^2 \rangle} \propto t^{1/2}$), and the
probability of finding the system very far away from the typical value
is exponentially low.  The eccentricity of the binary evolves
differently.  The probability of finding the binary with an
eccentricity larger than $e_c(t)$ only diminishes as a power law
(equation \ref{eqGofx}).  Physically, this reflects the probability
that the binary received a single large perturbation to that state.
The characteristic eccentricity, $\sim e_c(t)$ corresponds to the size
of the perturbation that occurs with a frequency of about $1/t$.  The
linear growth of $e_c(t)$ demonstrated by equation
\ref{eqDiffeoft} reveals that the eccentricity of the binary does not
reflect the accumulation of many small perturbations, but the single
largest perturbation occurring in its history.  This kind of random
walk is called a ``L\'{e}vy flight'' \citep{SZF95}.

\subsection{Inclination}
\label{secBEinc}

The same analysis applies to the changes in angular momentum of the
binary.  Since $|\Delta \vec i| \sim |\Delta \vec e|$, it follows that
$p(i') \sim p(e')$.  The evolution of inclination differs only in the
coefficients that depend on the geometrical configuration of the
encounter.  The calculation of the coefficients is described in the
Appendix.  The self-similar distribution shape is a function of the
dimensionless variable $i/i_c(t)$, where $i_c(t)$ is the
time-dependent characteristic inclination.  The following equation
describes the evolution of $i_c(t)$:

\begin{equation}
\label{eqDiffioft}
\dot i_c(t) = - i_c(t)/\tau_{d,i}+ \langle C_i \rangle G \rho T_{\rm orb}/2
\end{equation}

\noindent
where we have used $\tau_{d,i}$ to distinguish the timescale at which
the inclination of the binary is damped, and $ \langle C_i \rangle =
0.75$, the average of the angular terms in equation \ref{eqDeltah}.
The inclination is always measured relative to the orbital plane at
$t=0$. The distribution given by equation \ref{eqGofx} then describes
the probability of the binary being inclined by $i=x~ i_c(t)$ relative
to its original orbital plane.

\section{A Spectrum of Colliding Perturbers}
\label{secMassSpec}

For many physical applications we must consider a range of perturbing
masses and velocities and the effects of collisions onto the binary.
In the single mass case discussed in section
\ref{secBEecc}, the interaction frequency $p(e')$ is set by the
likelihood that the binary encounters a perturber at the impact parameter that
causes such a change of $e'$.  For perturbers that have different
masses, the chance of experiencing a perturbation of magnitude $e'$
depends on the combined likelihood that the perturber has the required
impact parameter and the required mass to excite such a change.

To extend our analysis we set up several pieces of notation.
Assuming that the mass and velocity distributions are independent, we 
consider ${\cal F}(m_p,v_p) = {\cal F}_v(v_p) {\cal F}_m(m_p)$.  We
restrict our analysis to velocity distributions with a characteristic
value, $v_0$, such as a Gaussian distribution.  We consider systems
with differential mass spectra characterized by a power law: ${\cal
  F}_m(m_p) \propto m_p^{-\gamma}$, valid from a minimum mass $m_{\rm
  min}$ to a maximum $m_{\rm max}$.  These functions are consistent
with conditions in the Kuiper belt, where a power law mass spectrum
and roughly Gaussian velocity spectrum are observed \citep{LJ02}.
We define the differential mass spectrum by 

\begin{equation}
\label{eqMassSpec}
{\cal F}_m(m_p)=(n_0 (\gamma-1)/m_0) (m_0/m_p)^{\gamma}, 
\end{equation}

\noindent
where $n_0$ is the number density of bodies larger than mass $m_0$.
In the literature the differential size spectrum of Kuiper belt
objects is characterized as a power law in
radius with index $q$; this is related to our index by $\gamma =
(q+2)/3$.  In this section we discuss the $p(e')$ and $p(i')$ that
result from several values of $\gamma$.

\subsection{$\gamma<2$}
\label{secShallow}

The total mass density of perturbers for $\gamma < 2$ is dominated by
the perturbers with the largest mass, $m_{\rm max}$.  While
perturbations of size $e'$ are excited by all of the perturbers, the
most likely perturber to cause a perturbation of this strength is the
largest mass perturber.  The dynamics of the binary are then the same as described
in section \ref{secBEecc} with $m_p=m_{\rm max}$.  The power law of
$p(e') \propto e'^{-3}$, based on distant encounters, is valid up to
the eccentricity excited by a perturber of mass $m_{\rm max}$
interacting at a $b \sim r_b$, or for $e' \ll (m_{\rm
  max}/m_b)(v_b/v_p)$ (equation \ref{eqDeltae}).  It is necessary only
to know the total mass density $\rho$ of the perturbing swarm in order
to calculate the excitation frequency in this scenario, given by
equation \ref{eqPofE}.

\subsection{$\gamma=2$}
\label{secEq2}

The power law $\gamma=2$ describes a special mass distribution where
the frequency of encountering the few large perturbers at large impact
parameters is the same as encountering the more abundant smaller perturbers at
smaller impact parameters.  Thus each logarithmic interval in impact
parameter contributes the same amount to the frequency of perturbations by 
$e'$, $p(e')$.
The upper limit of impact parameters that can contribute to
excitations of a given $e'$, however, is given by the maximum mass
perturber.  The total range of contributing impact parameters then
diminishes as $e'$ approaches the eccentricity caused by the largest
perturber interacting with $b \sim r_b$, $e'_{\rm max}\equiv (m_{\rm
  max}/m_b)(v_b/v_0)$.  Mathematically this behavior is determined by
the integral of equation \ref{eqRofe}, which yields an excitation
frequency of:

\begin{equation}
\label{eqRdistanteq2}
p(e') = \frac{G n_0 m_0 T_{\rm orb}}{e'^3} 
\frac{\log \left( 2.1 (e'_{\rm max}/e') \right)\langle C_e \rangle}
{4 \pi},
\end{equation}

\noindent
for $e' \ll e'_{\rm max}$.  The
equivalent formula for the inclination excitations is:

\begin{equation}
\label{eqRIdistanteq2}
p(i') = 
\frac{G n_0 m_0 
T_{\rm orb}}{i'^3} \frac{\log \left((e'_{\rm max}/i') \right)\langle C_i \rangle}{4 \pi}.
\end{equation}

\noindent
For the smallest $e'$ and $i'$, the entire range of perturbing masses
contributes to the interaction frequency.  This occurs for excitations of the order
$(m_{\rm min}/m_b)(v_b/v_0)$, below which the perturbation frequency is given by 
equation \ref{eqPofE}.

\subsection{$2<\gamma<3$}
\label{secInt}

The mass density of the perturbers when $2<\gamma<3$ is
dominated by perturbers of the smallest mass, $m_{\rm min}$.  Distant
encounters by perturbers with this mass produce very small
perturbations; for very low $e'$ then, $p(e') \propto e'^{-3}$, given
by the simple model of section \ref{secBEecc}.  The upper limit of
$e'$ caused by these perturbers interacting with impact parameters $b
\sim r_b$ is $e' \sim (m_{\rm min}/m_b)(v_b/v_p)$.  

Perturbers with
$m_{\rm min}$ cause eccentricity changes larger than this via close
encounters, but these encounters are less frequent than
interactions with perturbers of a higher mass and an impact parameter
of order $r_b$.  Perturbations with a strength $(m_{\rm
  min}/m_b)(v_b/v_p) \gg e' \gg (m_{\rm max}/m_b)(v_b/v_p)$ are most
often excited by perturbers with impact parameters of $\sim r_b$ and
masses $m \sim e' (v_p/v_b) m_b$.  In other words the frequency of
perturbations is directly proportional to the slope and normalization
of the mass spectrum.

In this case, the functions $p(e')$ and $p(i')$ cannot be determined
using the simplifications to equation \ref{eqDeltaegeneral} afforded
by very small or very large impact parameters.  In general, the
perturbation frequency for a mass spectrum of $2<\gamma<3$ follows the
power law $p(e') \propto e'^{-(\gamma+1)}$.  As an example we present
the perturbation frequency for $\gamma=25/12$.  This corresponds to
$q=4.25$, the best fit to observations of the Kuiper belt size
distribution presented by \citet{FKH08}.  We calculate from equation
\ref{eqRofe},

\begin{equation}
\label{eqPintermediate}
p(e')=2.6 \frac{G n_0 m_0 T_{\rm orb}}{e'^{37/12}} 
\left( \frac{m_0}{m_b} \frac{v_b}{v_0} \right)^{1/12}.
\end{equation}

\noindent
It is simple to understand the relationship between equations
\ref{eqPofE} and \ref{eqPintermediate} with the following argument.
A perturbation of size $e'$ that occurs via an interaction at a distance
$r_b$ requires a perturber of mass about $m' \sim e' (v_0/v_b) m_b$.
If we interpret the total density in equation \ref{eqPofE} as only
the density in bodies around $m'$, then $\rho' \sim m' {\cal F}_m(m') \sim
n_o (m_0/m')^{\gamma-1}$, and we recover the scaling of equation
\ref{eqPintermediate}.

The integral over
$b$ and the angular variables of equation \ref{eqDeltaigeneral} yield
a different coefficient for the perturbations to inclination:

\begin{equation}
\label{eqPofIintermediate}
p(i')= \frac{G n_0 m_0 T_{\rm orb}}{i'^{37/12}} 
 \left( \frac{m_0}{m_b} \frac{v_b}{v_0} \right)^{1/12}.
\end{equation}

\noindent
We relegate to the appendix the details of the integrals that 
produce the coefficients of equations \ref{eqPintermediate} and 
\ref{eqPofIintermediate}.



\subsection{Collisional Perturbations}
\label{secColRate}

The integral of equation \ref{eqRofe} over impact parameters from 0 to
$r_j$ produces the frequency of perturbations to the binary by
collisions on member $j$.  Since the size of the impulse from a
collision does not depend on the impact parameter, it is the mass of the
perturber that dictates the size of the eccentricity perturbation.
Accordingly, the frequency of perturbations as a function of $e'$
reflects the frequency of collisions as a function of $m_p$.  The frequency
of collisional perturbations does not depend on $m_{\rm max}$ or
$m_{\rm min}$ regardless of the slope.  However, the limits of the mass 
distribution specify the lowest and highest perturbations achievable 
via collisions: $\chi (m_{\rm min}/m_j)(v_0/v_b) \leq e' \leq \chi (m_{\rm max}/m_j)(v_0/v_b)$.
In this range of $e'$, for any value of $\gamma$, the
perturbation frequency due to collisions is

\begin{equation}
\label{eqPcol}
p(e') = \frac{G n_{0} m_b T_{\rm orb} }{e'^{\gamma+1}}
\left (\chi \frac{m_0}{m_j} \right)^{\gamma-1}
\left(\frac{v_0}{v_b}\right)^{\gamma} \left(\frac{r_j}{r_b}\right)^2
V_{\gamma} \frac{ (\gamma-1) \langle D_e^{\gamma-1}\rangle}{2 \pi},
\end{equation}

\noindent
where $\langle D_e^{\gamma-1} \rangle$ is the average of the angular
dependence of $\Delta \vec e$ from collisions to the power of
$\gamma-1$, and $V_\gamma \equiv v_0^{-\gamma} \int v_p^{\gamma+2}
{\cal F}_v(v_p) dv_p$.  If ${\cal F}_v(v_p)$ is proportional to a
delta function, $\delta(v_p-v_0)$, then $V_{\gamma}=1$ for all
$\gamma$.  If the velocity spectrum were Gaussian, such that ${\cal
  F}_v(v_p) \propto \exp (-(v_p/v_0)^2)$, then $V_\gamma=2
\Gamma((3+\gamma)/2)/\sqrt{\pi}$.  
The frequency of perturbations to the relative inclination by
collisions is the same as equation \ref{eqPcol}, replacing the
integrated coefficient $\langle D_e^{\gamma-1} \rangle$ with the
appropriate calculation made from the coefficients of $|\Delta \vec
i|$.

Although we use $r_j$ to represent either member of the binary, it is
clear from equation \ref{eqPcol} that the collisions onto the smallest
body have the largest effect on the orbit.  The ratio of the
perturbation frequency through collisions, $p(e')_{\rm collisions}$
(equation \ref{eqPcol}) to the frequency of gravitational scatterings,
$p(e')_{\rm gravity}$ (equation \ref{eqPintermediate}), is, for mass
distributions of $2 < \gamma <3$,

\begin{equation}
\label{eqRatio}
\frac{p(e')_{\rm collisions}}{p(e')_{\rm gravity}}
= 0.03 \left( \frac{r_j}{r_b} \right)^2
\left[\chi \frac{m_b}{m_j} \left( \frac{v_0}{v_b} \right)^2 \right]^{\gamma-1},
\end{equation}

\noindent
where we have evaluated the coefficients for $\gamma=25/12$.  The choice 
of $\gamma$ does not change these coefficients dramatically.





\subsection{Eccentricity Distributions}


The distribution given by equations \ref{eqGofx} and \ref{eqDiffeoft}
were derived in the context of $p(e') \propto e'^{-3}$.  As long as
$p(e')$ follows a power law with $e'$, we can write a self-similar
distribution function $f(e,t)$.  We write a generic function,
$p(e')=P_0 e'^{-(1+\eta)}$, to account for the different slopes caused
by different mass distributions (for $3 > \gamma > 2$, $\eta =
\gamma$; for $\gamma < 2$, $\eta=2$).  The derivation of the
distribution function proceeds analogously as in section
\ref{secBEecc}.  Equation \ref{eqFofe} becomes two equations: a
dimensionless integro-differential equation that specifies the shape,
and an ordinary differential equation to specify the evolution of the
eccentricity scale $e_c(t)$.  The general version of equation
\ref{eqDiffeoft} is:

\begin{equation}
\label{eqDiffeoftgeneral}
\dot e_c(t) = - e_c(t)/\tau_d + 2 \pi P_0 / e_c(t)^{\eta-2}. 
\end{equation}

\noindent 
In the limit of no eccentricity dissipation ($\tau_d \rightarrow
\infty$), equation \ref{eqDiffeoftgeneral} shows that $e_c(t) \propto
t^{1/(\eta-1)}$. For all of the $p(e')$ discussed in section \ref{secMassSpec},
the growth of $e_c(t)$ is always faster than $t^{1/2}$.

The shape of the distribution function is determined through a Fourier
transform of the general version of equation \ref{eqFofg}.  For slopes
of $1<\eta<3$, $g(x)=\int \cos (\vec k \cdot \vec x) \exp(-|\vec
k|^{\eta-1})d^2 \vec k $ \citep{S99,CSS07}.  While there is only a
closed form solution for $\eta=2$, given by equation \ref{eqGofx}, all
of these functions are flat at low $x$ and fall off like
$x^{-(\eta+1)}$.  In fact, it is easy to show from equation
\ref{eqFofe} that the high $e$ tail is given by

\begin{equation}
\label{eqHighTail}
f(e \gg e_c(t)) = p(e) t / (\gamma-1),
\end{equation}  

\noindent
when $t \ll \tau_d$.  For equilibrium distributions where $\dot
e_c(t)=0$, $t$ is replaced with
$\tau_d$, the timescale for the dissipation.

When $p(e') \propto e'^{-4}$ or steeper, the accumulation of the
smallest perturbations over time is more effective at raising the
eccentricity of the binary than single large perturbations.  In this
case, the evolution of the eccentricity follows standard Brownian
motion, where the distribution function is a Gaussian, and $e_c(t)
\propto t^{1/2}$.  

\section{Kuiper Belt Binaries}
\label{secKBB}

In this section we compute $e_c(t)$ and $i_c(t)$ for several Kuiper
belt binaries.  The ``binary'' of section \ref{secSingleEnc} now refers to 
a bound pair of Kuiper belt objects, and the ``perturbers'' are all of the 
other members of the Kuiper belt.

For the highest mass KBOs, the size spectrum is well determined to be
a power law with an index slightly greater than $q=4$.  The lowest
mass bodies, of about 30 km in radius, are less frequent than
predicted by a single power law, however the parameters of a more
general model are still under investigation
\citep{TB01,LJ02,PS05,FKH08,FH08}.  For this section we use the best
fit of a single power law model to the high mass part of the spectrum
provided by \citet{FKH08}, who find $q=4.25$ and a number density of 1
body per square degree brighter than magnitude 23.4.  We assume an
average distance of 40 AU to the Kuiper belt and a depth of 20 AU to
find a volumetric number density $n_0=3 \times 10^{-41} ~{\rm
  cm^{-3}}$.  To convert the magnitudes of the objects to physical
sizes, we assume a constant geometric albedo of 0.04, a constant
physical density of 1 g ${\rm cm^{-3}}$, and take the R-band apparent
magnitude of the Sun to be -27.6.  We find that the magnitude 23.4
corresponds to a mass $m_0=1.75\times 10^{21}~{\rm g}$, equivalent to
a radius of 75 km.  Most of the objects found between 30-50 AU are
inclined by about 5-15 degrees relative to the plane of the solar
system, and have heliocentric eccentricities of 0.1-0.2.

\subsection{Perturbations by a Disk}
\label{secDisk}

Our analysis so far has treated the perturbing bodies as unbound
objects moving relative to the binary with a constant velocity.  When
the perturbers are part of a disk orbiting the central star, the
orbital elements of the disk set the parameters of the perturbation
frequencies we calculate in section \ref{secBE}.

The relative velocity between KBOs, when they interact, is set by the
size of their eccentricities and inclinations, $v_p \sim e_{H} a
\Omega_H$, where the subscript ``H'' denotes a heliocentric orbital
quantity.  We assume a constant perturbing velocity with $v_p = 1~{\rm
  km/s}$, which corresponds to the typical heliocentric eccentricities
and inclinations of KBOs.  We assume that these encounters occur
isotropically in the frame of a binary, however this is not accurate.
A more detailed calculation of the angular distribution of relative
velocities will only affect the coefficients of the perturbations.
The disk does not specify a special direction for the perturbation vector
$\Delta \vec e$, so the perturbing frequency and the distribution
function retain their axisymmetry.  The influence of the central 
star on the binary and the perturbers adds
another constraint to our assumption of impulsive encounters: the
timescale for an interaction must be shorter than the orbital period
around the star: $b/v_p \ll 1/\Omega_H$, or equivalently, $b \ll e_H
a$.  This guarantees that the relative velocity is constant 
during the interaction.

If the orbit of the binary is much different than the typical KBO
orbit, there are several modifications to perturbation frequencies
experienced by the binary.  One modification is due to the finite
height of the disk of perturbers.  This height is set by their
inclinations around the central star; for the Kuiper belt we refer to
the average inclination as $\langle i \rangle_{KB}$.  A binary with
heliocentric inclination $i_{\rm CoM} \ll \langle i \rangle_{KB}$
never travels above or below the perturbing disk height and therefore
experiences the maximal frequency of perturbations.  If $i_{\rm CoM} \gg \langle
i \rangle_{KB}$, the binary spends most of its orbit outside of the
perturbing swarm.  The frequency of perturbations to such a binary is
reduced by the fraction of the time the binary leaves the disk,
proportional to $\langle i \rangle_{KB}/i_{\rm CoM}$.  The
eccentricity of the binary in the disk reduces the effective density
of perturbers in a similar manner if the epicycle of the binary
carries it outside of the region populated by perturbers.

If the heliocentric eccentricity or inclination of the binary is much
greater than the typical values for the Kuiper belt, the relative
velocity between the binary and a perturber is primarily due to 
the non-circular heliocentric motion of the binary.
Gravitational interactions depend weakly on $v_0$ so their frequency
does not change much in this case.  Perturbations by collisions,
however, become more important if $v_0$ is increased due to this
effect (equation \ref{eqRatio}).

\subsection{Pluto et. al.}
\label{secPluto}

Pluto is the second largest known Kuiper belt object, with a radius of
about 1100 km.  It has a semi-major axis of 39.5 AU and its orbit is
inclined relative to the ecliptic by $17^{\circ}$.  Its largest
satellite, Charon, contains about one tenth of the total mass of the
system.  Recent observations have revealed two smaller satellites, Nix
and Hydra \citep{W06}.  These satellites have small eccentricities and
are roughly co-planar with Charon.  Numerical simulations of
collisions between similarly sized objects by \citet{C05} produce
binaries with orbits similar to Pluto and Charon.  The circularity and
co-planarity of Nix and Hydra lend additional weight to a collisional
origin of the system.

The triple system of Pluto and its moons is a valuable test case for
the dynamics we have presented.  For an isolated binary it is
impossible to know the initial orbital plane.  The relative
inclinations of the moons of Pluto can be measured directly assuming
their formation was co-planar.  Furthermore, the perturbing swarm for
all three Pluto-moon pairs is the same.  A major issue in comparing
our analytic calculations to the observations is that the large mass
ratio of Charon to Pluto causes significant non-Keplerian effects in
the orbits of the outer satellites.  We first re-examine the published
observational model of their orbits to separate the relevant motion
of the outer satellites from the forced motion due to Charon.  We then
compare the resulting eccentricity with our predicted values.

\subsubsection{Orbital Model of Tholen et al}

A model of the observations of the Pluto system
has been presented by \citet{TBGE07}, who fit the parameters of a
four-body numerical integration such that the simulation agrees with
the observations.  Such work is necessary, as it has been 
shown that the observations cannot be consistently modeled by 
three non-interacting two-body orbits \citep{W06}.

The model of \citet{TBGE07} presents a full set of osculating elements
describing the orbits of Charon, Nix, and Hydra.  The orbit of Charon
is virtually unaffected by Nix and Hydra; \citet{TBGE07} measure the
eccentricity of Charon to be $3.48 \pm 0.04 \times 10^{-3}$, and the 
period of its orbit is $6.387$ days. Since
the combined potential of Charon and Pluto is significantly
non-Keplerian, the elements of Nix and Hydra vary significantly during
their orbits.  \citet{TBGE07} average the osculating semi-major axis
to find an orbital period for these satellites of 25.49 days and 38.73
days for Nix and Hydra respectively.  The osculating eccentricities of
Nix and Hydra both oscillate between zero and about 0.2; for each
satellite oscillations at the frequencies of its own orbit and
that of Charon are visible (their figure 4).  The orbital planes of the 
satellites relative to Charon's are tilted by 0.15 degrees for Nix and 
0.18 degrees for Hydra.  Each plane precesses relative to the plane of Charon, 
however the angle of the offset remains constant.

\subsubsection{A Different Interpretation}
\label{secInterpret}

\begin{figure}[t!]
\center
\includegraphics[angle=-90,width=0.8\columnwidth]{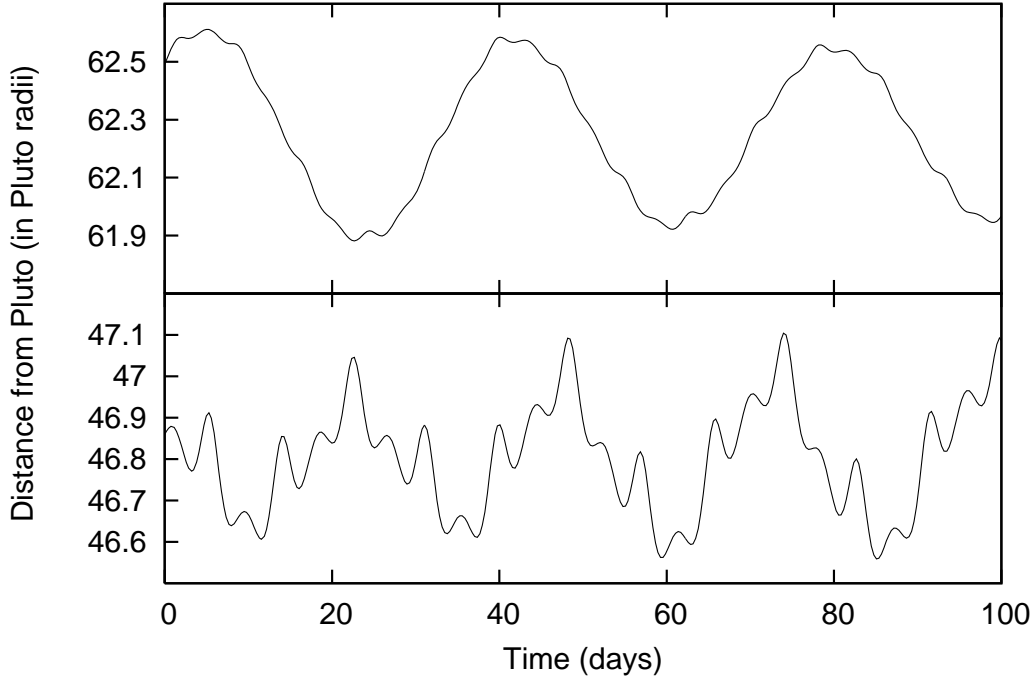}
\caption{The distance of Nix (lower panel) and Hydra (upper panel) from the Pluto-Charon barycenter, in units of Pluto radii, as a function of time, in 
an integration of the parameters found by \citet{TBGE07}.  Nix and Hydra are treated
as massless test particles.  The origin of the time coordinate is arbitrary.}
\label{figRNixHydra}
\end{figure}

For two body motion, the Keplerian elements are constant and indicate
the shape of the orbit in space.  Osculating elements that describe
motion in significantly non-Keplerian potentials, such as the combined
potential of Pluto and Charon, may vary on timescales shorter than the
orbital period of the satellite.  When this is true, relating the
osculating elements to the shape of the orbit can be misleading. 
The average value of the osculating eccentricity of Nix is
0.015 in the model of \citet{TBGE07}, however the motion of Nix
relative to Pluto never resembles an ellipse with such an
eccentricity.

We re-examine the model provided by \citet{TBGE07} by reproducing the
numerical integration based on the Pluto-centric positions and
velocities of Charon, Nix, and Hydra published in their table 1.  We
set the masses of Nix and Hydra to zero to eliminate their secular
interactions with each other.  Instead of examining the
osculating elements, we adopt the approach of \citet{LP06}
and characterize the orbits of Nix and Hydra based
on their position as a function of time from the Pluto-Charon
barycenter, plotted in figure \ref{figRNixHydra}.  The units of distance
are Pluto radii, defined as $R_P = 1147~{\rm km}$.  

Although short oscillations on the timescale of Charon are visible in
the top panel of figure \ref{figRNixHydra}, they are very small
compared to the oscillations that occur on the timescale of Hydra's
orbital period.  To parametrize Hydra's orbit we fit the function
$r_0(1+e \cos (\kappa_1 t + \omega_1))$ to the first 200 days of the
numerical model.  Because for a non-Keplerian potential the radial
epicyclic frequency differs from the orbital frequency, we calculate
the average angular frequency by fitting a straight line to the
angular position of Hydra as a function of time, $f(t) = \Omega_1
t+\lambda_0$.  The results are written in table \ref{tabFitResults}.
We interpret $e_1$ as the orbital degree of freedom in the combined
potential of Pluto and Charon that is analogous to the eccentricity of
a two-body orbit.

\begin{deluxetable}{lcccccccc}
\tablecaption{Best fit values to the epicyclic models of the radial motion
of Nix and Hydra.}
\tablehead{
\colhead{} & \colhead{$r_0/R_P$} & \colhead{$e_1$} 
& \colhead{$2 \pi/\kappa_1$}& \colhead{$e_2$} 
& \colhead{$2 \pi/\kappa_2$}&  \colhead{$e_3$} 
& \colhead{$2 \pi/\kappa_3$}& \colhead{$2 \pi/\Omega_1$} \\
\colhead{} & \colhead{} & \colhead{$\times 10^{-3}$} & \colhead{(days)} 
& \colhead{$\times 10^{-3}$} & \colhead{(days)} 
& \colhead{$\times 10^{-3}$} & \colhead{(days)} & \colhead{(days)}}
\startdata
Nix & 46.805(5) & 2.96(3) &  25.22(2) &
 1.25(3) &  8.599(8) &  1.38(3) & 4.298(1) & 24.8505(5) \\
Hydra &  62.237(1) &  5.595(2) & 38.535(15) & & & & & 38.20(1) \\
\enddata
\tablecomments{The motion of Nix is fit with three epicyclic 
terms, while the motion of Hydra is only fit with one.  The parenthesis 
indicate the 95 \% confidence level of the fit around the last digits.}
\label{tabFitResults}
\end{deluxetable}

The motion of Nix (bottom panel of figure \ref{figRNixHydra}) appears
more irregular than that of Hydra. We find the position of Nix to be
well-described by a model of three epicycles with different
frequencies: $r(t)=r_0(1+\sum_{k=1,2,3} e_k \cos (\kappa_k t +
\omega_k))$.  The best fit values are printed in table 1.  We
distinguish the cause of each epicycle by its period.  The combined
potential of Pluto and Charon oscillates with frequency of
$\Omega_{\rm Charon}-\Omega_{\rm Nix}$; motion being forced by this
potential should occur on integer multiples of this frequency.  
Using the numbers in table \ref{tabFitResults}, we see that 
$2 \pi/(\Omega_{\rm Nix}+\kappa_2) = 2\pi/(\Omega_{\rm Nix}+\kappa_3/2)=
6.39$ days.  The second and third epicycles in our fit correspond 
to motion at the first and second harmonic of Nix's relative 
orbital frequency.  We therefore interpret the first term, with a size of 
$e_1=3 \times 10^{-3}$ and a period close to Nix's orbital period,
as analogous to the two-body eccentricity.

We perform another integration of the best fit initial conditions from
\citet{TBGE07} to investigate the secular effects between Nix and
Hydra.  We use the best fit masses from \citet{TBGE07} for the two
outer satellites.  Since the motion of Hydra is dominated by a single
epicyclic frequency, the variation in the size of its epicycle is
apparent on the timescale of several years.  To determine the effect
of secular variations on Nix, we fit the same three-component
epicyclic model to five orbits at $t \sim 5$ years.  In the best-fit model to these
later orbits, the only difference compared to the model of table
\ref{tabFitResults} is in $e_1$, the epicycle with a frequency close
to Hydra's orbital frequency.  This is further confirmation that the
degree of freedom represented by $e_1$ is analogous to the two-body
eccentricity.  

\subsubsection{Theoretical Distribution}

To compute the distribution of eccentricities and inclinations
expected of Pluto's moons, we solve equation \ref{eqDiffeoftgeneral}
for each of the moons, given the interaction frequencies specified by
equations \ref{eqPintermediate} and \ref{eqPofIintermediate}.  The
only remaining parameters to evaluate are the damping timescales for
the eccentricity and inclinations of each satellite.  We use the standard 
formula for the damping of eccentricity due to the tidal 
force of the primary acting on a secondary that is in synchronous 
rotation \citep{YP81,MD99}:

\begin{equation}
\label{eqTidalDamping}
\tau_{d,2} = \frac{4}{63} Q_2 (1+{\tilde \mu_{2}}) \frac{m_2}{m_1}
\left( \frac{r_b}{r_2} \right)^5 \frac{1}{\Omega},
\end{equation}

\noindent
where $Q_2$ is the dissipation function of the secondary, and 
${\tilde \mu_2} = 19 \mu r_2 / (2 \rho G m_2)$ 
is its effective rigidity, a ratio between the 
material strength of the secondary and its self-gravity. 
The damping rate of
eccentricity due to tides of the primary acting on the secondary,
$\tau_{d,1}$, if the primary is also rotating synchronously with the
orbit of the satellite, is given by equation \ref{eqTidalDamping} with
the quantities specific to the primary switched with those of the
secondary and vice versa.

Pluto and Charon are known to be in a double-synchronous state of
rotation, where the spin period of each body is equal to the 6.4 day
orbital period.  In many binaries, only the spin of the secondary is
synchronous with the orbital period.  Tides on the primary then raise
the eccentricity.  Double-synchronous systems, however, experience
damping due to both the tides on the secondary and those on the
primary.  Assuming a water-ice composition for Pluto 
($\mu=4~\times 10^{10}~{\rm dynes~cm^{-2}}$), we calculate the
eccentricity damping timescale due to tides raised by Charon,
$\tau_{d,1}$ from equation \ref{eqTidalDamping} to be 5.1 Myrs.  The
shortest damping timescale due to tides from Pluto acting on Charon,
$\tau_{d,2}$ is found by assuming Charon is also made of water-ice; we
find in this case a timescale of 8.2 Myr.  The longest timescale
assumes a rocky composition ($\mu=6.5~\times 10^{11}~{\rm dynes~cm^{-2}}$); 
we find this corresponds to 133 Myr.  The
overall damping of the system is given by the sum of the damping
rates.  The short damping timescale of tides on Pluto prevents Charon
from contributing significantly to the combined effect of both tides,
reducing the importance of its composition.  The longest eccentricity
damping timescale that results from both tides is 4.9 Myr.  The
inclinations of the outer satellites relative to the Pluto-Charon
plane are also damped by tidal dissipation. For a circular synchronous
orbit the timescale for inclination damping is longer than the
timescale for eccentricity damping by a factor of $\sim i^{-2}$.  We
ignore the damping of inclinations in equation \ref{eqDiffioft} for
all three satellites.

As discussed in \citet{TBGE07} and section \ref{secInterpret}, secular
interactions between the satellites are visible in the long term
calculations of their orbits.  For the best-fit values of the masses
of Nix and Hydra, their eccentricities are modulated on the order of
$10 \%$ over timescales of years; we neglect these fluctuations for
this work.  It is more important in this model to determine whether
secular evolution can cause the eccentricity of Nix or Hydra dissipate
via Charon's orbit.

We use linear secular theory to describe the coupled evolution of the
eccentricity and longitude of periapse of each satellite \citep{MD99}.
We find that the undamped secular evolution agrees qualitatively with
the numerical orbit determinations.  We add a term to the differential
equations describing Charon's eccentricity that reduces it at a
constant timescale ($\dot e_{\rm Charon} = - e_{\rm Charon}/\tau_d$).
The frequencies of the oscillations of the eigenmodes of the solution
are practically unchanged by this term, however each eigenmode gains a
dissipative factor.  Quantitatively, only one eigenmode is damped on
timescales shorter than than 4.5 Gyr.  By integrating the damped
secular equations with different initial periapses, we determined that
the secular interactions do not cause substantial damping of Nix and
Hydra.

Equation \ref{eqRatio} gives the
frequency of perturbations due to collisions of perturbers onto each
moon relative to the frequency of perturbations caused by
gravitational scattering, equation \ref{eqPintermediate}.  For Charon,
the collisional perturbations increase $p(e)$ by only 2 percent.
Since Nix and Hydra are smaller, perturbations by collisions have a
greater relative effect; however it is only a 20 percent contribution
to the total perturbation frequency for Nix and 15 percent for Hydra.
We solve equation \ref{eqDiffeoftgeneral} to find $e_c(t)$ and
$i_c(t)$ for each of Pluto's moons. 

For Charon we find $e_c=2.6 \times 10^{-6}$, and $i_c=0.029^{\circ}$.
This value of $e_c$ corresponds to the most likely perturbation during
a damping timescale of 4.9 Myr, and is much smaller than the
observed value of $3.5 \times 10^{-3}$ \citep{TBGE07}.  Using equation
\ref{eqHighTail}, we calculate that given this value of $e_c$, the
probability of Charon's eccentricity being as high as its observed
value is 0.2 percent.

For Nix we calculate $e_c(4.5 ~{\rm Gyr})=4.8\times 10^{-3}$ and
$i_c(4.5 ~{\rm Gyr})=0.1^{\circ}$, and for Hydra, $7.1 \times 10^{-3}$
and $0.15^{\circ}$ respectively.  The distributions specified by these
values are quite consistent with the free eccentricity we determine in
table \ref{tabFitResults}.

\subsection{Other Interesting KBOs}
\label{secOthers}

Two other Kuiper belt objects have satellites on low eccentricity
orbits: \el, and Eris.  Along with Pluto these are three of the four
most massive KBOs known, all with radii of about 1000 km.
\el has two known satellites.  The largest has a 50 day orbit and a
measured orbital eccentricity of $0.050 \pm 0.003$ \citep{BBR05}.  An
additional smaller satellite orbits \el~ with a period of about 35
days \citep{BVB06}.  The orbital parameters of the inner satellite are
unconstrained, however the relative inclination between the two is
about $40^{\circ}$.  The masses of the satellites are negligible compared
to the mass of \el.  The heliocentric inclination of the system is
$28^{\circ}$.

\citet{BBR05} argue that if the tidal response of \el~ and its large
satellite are fluid-like, tidal interactions should damp their
eccentricity on a timescale of about 300 Myr.  With these parameters
we use equation \ref{eqDiffeoftgeneral} 
to calculate an equilibrium $e_c=4.3 \times 10^{-4}$.  The
distribution with this eccentricity scale predicts an observed
eccentricity of 0.05 at a probability of three percent.  However, for
smaller bodies, internal elastic forces dominate the tidal
deformation of their shape; it is more reasonable to assume that the
tidal response of the satellite is characterized by its material
strength.  Then, the tides raised on the primary have the greatest effect
and the eccentricity of the system grows on the same timescale as the
growth of the semi-major axis.  Forced eccentricity growth and an
evolving orbital period can be incorporated into equation
\ref{eqDiffeoftgeneral}.  However, these corrections are only an order
unity correction since the growth timescale, by definition, is
comparable to the age of the system.  Assuming $T_{\rm orb}$ is fixed
and ignoring the eccentricity growth, we calculate $e_c(4.5~{\rm
  Gyr})=0.0052$.  The 95 percent confidence interval around this $e_c$ is 
0.001-0.2; the observed eccentricity of \el is within this range.

The dwarf planet Eris is orbited by the satellite Dysnomia.
Observations have shown an upper limit to their eccentricity of 0.013
\citep{BVB06}.  The system has a 15 day orbital period, and orbits the
sun at a semi-major axis of 67.7 AU with an eccentricity of 0.44 and a
heliocentric inclination of $44^{\circ}$.  In addition to the
reduction in effective perturbing density caused by the inclination,
the high eccentricity reduces the effective perturber density by an
additional factor of 0.09.  The semi-major axis of the binary is
consistent with 4.5 Gyr of tidal evolution away from an initially very
close orbit; if the tidal response of the secondary that of a
strength-less fluid, then its eccentricity is damped on a timescale of
50 Myr.  These parameters yield an $e_c=2.2\times 10^{-6}$.  However,
if the material strength of the secondary is stronger than its own
self-gravity, then the tides raised on the primary cause the
eccentricity of the satellite to grow.  In this case the relevant
timescale is the age of the system, and we find that $e_c(4.5~{\rm
  Gyr})=1.0\times 10^{-4}$.  Both values are below the observed upper
limit.

In addition to the high mass ratio and low eccentricity Kuiper belt
binaries, there are other known binaries of almost equal mass on
moderately eccentric orbits.  The binary $1998~{\rm WW}_{31}$ is an
example of such an object: both members have a radius of about 50 km,
an orbital period of 574 days, and a mutual eccentricity is 0.817
\citep{VPG02}.  Even though our analysis is derived in the
low eccentricity limit, we can use equation \ref{eqDiffeoftgeneral}
to estimate approximately the eccentricity expected from impulsive
encounters; we find $e_c(4.5~{\rm Gyr}) = 0.31$.  This moderate
characteristic eccentricity is consistent with the high observed
value.  Other binaries with orbital periods on the order of a year
will have acquired large eccentricities through their interactions
with the other Kuiper belt objects.

\section{Other Binary Systems}
\label{secApp}

Our analysis holds for any two-body orbit perturbed isotropically in
the impulsive limit.  As binary orbits are prevalent in astrophysics,
we briefly discuss several other examples.

The asteroid belt harbors many binaries with well determined
eccentricities.  The mass spectrum of the asteroid belt, however, is
much shallower than that of the Kuiper belt: the largest asteroid,
Ceres, contains a third of the total mass of all asteroids.  A binary
asteroid is then perturbed mostly by the largest objects that it
encounters.  To calculate $p(e')$ accurately, it is necessary to model
the neighborhood of that binary.  The asteroid belt is also
collisionally active so its binaries may not be coeval with the whole
solar system. We postpone a detailed analysis of the binary asteroid
population for a future work.

A well-measured class of binaries outside the solar system are
millisecond pulsars with white dwarf companions.  The tidal damping
between the pulsar and its companion in the phase before the companion
becomes a white dwarf is very short, indicating that during this phase
the eccentricity of the binary should be smaller than the observed
values of around $10^{-4}-10^{-5}$ \citep{S04}.  To explain the
observations, \citet{P92} presents the following model.  As the
companion star becomes a white dwarf, random fluctuations in the
atmosphere of the star cause irregular motion in the orbit of the
binary.  These motions are reflected by a small eccentricity that
remains since the tidal interactions between the white dwarf and the
neutron star cannot damp the system.  The model of \citet{P92} produces
eccentricities for these systems that match the observations well.

These binaries are perturbed by encounters with other stars in the
galaxy; we can calculate the contribution to their eccentricities by
the distant stellar interactions.  The perturbation of these systems
by other stars falls into the simple regime of only distant
interactions described in section \ref{secBEecc}.  A typical
volumetric mass density for field stars is $0.1 {\rm M_{\odot}~
  pc^{-3}}$ \citep{HF00}.  Given this density, we calculate the
characteristic eccentricity of these systems to be

\begin{equation}
\label{eqEcpulsars}
e_c(t)= 1.2 \times 10^{-9}  \left( \frac{T_{\rm orb}}{1~{\rm day}}\right)
\left( \frac{t}{1~{\rm Gyr}} \right)
\left( \frac{\rho}{0.1 M_{\odot}~{\rm pc}^{-3}}\right).
\end{equation}

\noindent
Typical orbital periods are between 1 and 10 days, and the ages of
these systems are on the order of Gyrs.  We find then that $e_c(t)$ is
several orders of magnitude lower than the observed eccentricities.
\citet{P92} also concludes that the perturbations from other stars
cannot be responsible for the eccentricities of the binary pulsars.
Since we have calculated the distribution, however, we can estimate
more accurately the likelihood of achieving these eccentricities
by only distant stellar perturbations: less than 0.1 percent.

Globular clusters can have densities many orders of magnitudes higher than
the average galactic density, such that distant perturbations to the 
binaries may be important.  However, in a cluster the 
interactions between a binary and a star are not
typically in the impulsive interaction regime.  Instead the orbits of
the perturbers are affected by the gravity of the binary, and the
interactions occur over several orbital periods.  Analytic work on the
eccentricity perturbations in this regime has been performed by
\citet{RH95} and \citet{HR96}.

The characteristic eccentricity caused by distant stellar passages on
the orbits of extra-solar planets is also given by equation
\ref{eqEcpulsars}.  These eccentricities are too low to be
reflected in the current sample of known extra-solar planets.  As with
the pulsar binaries, the distant interactions may play a role in
setting the eccentricity distribution of long period planets found in a dense
stellar cluster.  For most extra-solar planets however, planet-disk 
interactions \citep{GS03} or planet-planet scatterings \citep{RF96}
are probably the source of their eccentricity.

\section{Conclusions}
\label{secConclusions}

We have calculated the effects of impulsive perturbations and
collisions on a nearly circular Keplerian orbit.  If the swarm of
perturbers encounter the binary isotropically in space, we can write a
distribution function that describes the probability density for the
binary to have a given eccentricity or inclination relative to its
initial plane. The growth rate of the binary's likeliest eccentricity 
and inclination depends on the mass spectrum of the perturbers.  For 
shallow mass distributions ($q<4$) it is the distant encounters that 
set the binary's eccentricity and only the total mass density of perturbers
is important to the evolution of the binary.  For steeper mass distributions of 
$q=4-7$, it is the interactions at about the semi-major axis of the binary 
that dominate the frequency of perturbations.  Only the normalization and 
slope of the 
mass spectrum set the distribution of eccentricities in this regime.

The assumptions of this model are valid in the Kuiper belt.  Our
calculations match the observations of Nix and Hydra very well.  For
Eris and \el, the observations lie within the 95 percent confidence
intervals of the distributions we calculate, assuming the tidal
response of the secondaries is dominated by material strength.  For
Charon our theory is consistent with the numerical simulations of
\citet{SBL03}, predicting an eccentricity about 3 order of magnitudes
smaller than observed.  However, our analysis alleviates their need
for numerical simulations as well as predicts the entire distribution
of the eccentricity.  The distributions measured by \citet{SBL03} are
not all correct as their model includes only impact parameters out to
twice the semi-major axis.  In their simulations where $q=3.5$ and 4.0
this excludes the impacts that are most relevant over an eccentricity
damping timescale.  Our results show that for $q=3.5$ the interactions
that dominate Charon's eccentricity are Pluto-sized perturbers
interacting at about 200 times the semi-major axis!


Even without eccentricity dissipation through tides, perturbations
from other Kuiper belt objects are too weak to excite eccentricities
of order 1 or inclination changes of order a radian for binaries that
have orbital periods of a few days or weeks.  It is not likely that
the orbital planes of the close binaries have been affected
significantly by other Kuiper belt objects given our current
understanding of the history of the Kuiper belt.  It falls on theories
of binary formation to explain the distribution of orbital
inclinations relative to the ecliptic for close binaries.  Since
$e_c(t)$ grows faster for binaries with large orbital periods, it is
plausible that the smaller wide binaries ($1998 {\rm WW}_{33}$ for
example) have been brought to large eccentricities and inclinations by
interacting with the rest of the Kuiper belt.

When many binaries share the same perturbing swarm, such as in the
Kuiper belt, we can use the eccentricities of all the binaries to
probe the properties of the entire system.  For example, if the mass
spectrum is steeper than $q=4$, the distribution of 
eccentricity is directly related to the slope and
normalization of the mass spectrum. 
Conversely, the observed eccentricity can be used to place limits on
the damping timescale of a binary and therefore the rigidity of those
bodies.  The small sample of Kuiper belt binaries with well measured
eccentricities limits the current effectiveness of such a calculation.
However, the Pan-STARRS project plans to detect around 20000 more
members of the Kuiper belt \citep{KAB02}; from these the number of
orbit-determined Kuiper belt binaries will surely increase.

The distribution we describe with equation \ref{eqGofx} is a special
case of a L\'{e}vy distribution \citep{S99}.  This class
of functions arise in the generalization of the central limit theorem
to variables distributed with an infinite second moment.
Alternatively, these functions can be characterized by the properties
of the L\'{e}vy flight they describe.  For the eccentricity of the
binaries discussed in this work, the frequency of a step is inversely
proportional to a power of its size that depends on the
mass spectrum of perturbers.  It follows that the largest single step
dominates the growth from accumulated smaller steps, causing, in the
absence of damping, the typical eccentricity to grow faster than in a
normal diffusive random walk.  The slope of the distribution of
excitations dictates the shape of the distribution.  This explains the
coincidence of the distribution we derive in this work being exactly
that of the distribution of eccentricity of protoplanets in a
shear-dominated planetesimal disk, where the probability of changing the
eccentricity of a protoplanet is inversely proportional to the size of
that change \citep{CS06,CSS07}.

The authors thank Dmitri Uzdensky and Scott Tremaine for valuable
discussions.  R.S. is a Packard Fellow and an Alfred P. Sloan Fellow.
This research was partially supported by the ERC.

\appendix

To calculate the excitation rates presented in sections \ref{secBE}
and \ref{secMassSpec}, it is necessary to integrate over all possible
configurations of angles $\vec b$ and $\vec v_p$ relative to $\vec
r_b$ and $\vec v_b$.  In this appendix we clarify the relation between
the coefficients and equations \ref{eqDeltaegeneral} through
\ref{eqDeltah}.

We choose spherical polar
coordinates for $\vec b$ and $\vec v_p$ to integrate 
equation \ref{eqRofe}.  This requires a
polar and azimuthal angle for $\vec b$, $\theta_b$ and $\phi_b$, and a
polar and azimuthal angle for $\vec v_p$, $\theta_v$ and $\phi_v$.
By defining $\theta_v$ relative to $\vec b$, the requirement that
$\vec b$ and $\vec v_p$ be perpendicular fixes $\theta_v = \pi/2$.

The magnitude of the perturbation only depends on the vectors
$\vec b$ and $\vec v_p$ relative to $\hat r_b$ and $\hat v_b$, so we use
these vectors and their cross product, $\hat n$ to describe the
components of $\hat b$: $\hat b = b_r \hat r_b + b_v \hat v_b + b_n
\hat n $.  The components are related to $\theta_b$ and $\phi_b$ in
the typical way: $b_r=\cos \phi_b \sin \theta_b$, $b_v=\sin \phi_b
\sin \theta_b$, and $b_n= \cos \theta_b$.  
We define the components
of $\vec v_p$ relative to the same unit vectors.
The angle $\phi_v$ describes the direction of $\vec v_p$ in
the plane given by $\hat b$; the components of $\vec v_p$ follow
from a rotation of this plane to align with $\hat n$.
We find the relations:

\begin{eqnarray}
\label{eqVconversion}
v_r & =& b_n\cos \phi_v 
- b_v ( b_r \sin \phi_v - b_v \cos \phi_v) / (1+b_n), \nonumber \\
v_v & =& b_n \sin \phi_v +  b_r (b_r \sin \phi_v - b_v \cos \phi_v) 
/ (1+ b_n),\\
v_n& =& - b_r \cos \phi_v - b_v \sin \phi_v. \nonumber
\end{eqnarray}

The coefficient from equations \ref{eqPofE} and \ref{eqRdistanteq2},
$\langle C_e \rangle$, is defined to be the integral of   
$|\Delta \vec e|/(8 \pi^2 (m_p/m_b)(v_b/v_p)(r_b/b)^2)$ as given by
equation \ref{eqDeltae}:

\begin{equation}
\label{eqAvgCe}
\langle C_e \rangle = 
\frac{1}{4 \pi^2} \int_0^{2 \pi} \int_0^{2 \pi} \int_{0}^{\pi} 
\left[ (4 b_r b_v +2 v_r v_v)^2+(1-v_r^2-2 b_r)^2
\right]^{1/2} \sin \theta_b d\theta_b d\phi_b d\phi_v =1.89
\end{equation}

We similarly define $\langle C_i \rangle$ from equation \ref{eqDeltah}:

\begin{equation}
\label{eqAvgCi}
\langle C_i \rangle = 
\frac{1}{4 \pi^2} \int_0^{2 \pi} \int_0^{2 \pi} \int_{0}^{\pi} 
| 2 b_r b_n+ v_r v_n | \sin \theta_b 
 d\theta_b d\phi_b d\phi_v = 0.75.
\end{equation}

To calculate the coefficients used in the collisional excitation rate,
equation \ref{eqPcol}, we use the $|\Delta \vec e|$ discussed in 
section \ref{secCollisions}.

\begin{equation}
\label{eqAvgDe}
\langle D_e^{\gamma-1} \rangle = 
\frac{1}{4 \pi^2}  \int_0^{2 \pi} \int_0^{2 \pi} \int_{0}^{\pi}
(4 v_v^2+v_r^2)^{(\gamma-1)/2}
\sin \theta_b  d\theta_b d\phi_b d\phi_v
\end{equation}

\noindent
For $\gamma=2$, the integral has a closed form solution of 
$\langle D_e \rangle = E(-3)$, the complete Elliptic integral.
For the inclination,

\begin{equation}
\label{eqAvgDi}
\langle D_i^{\gamma-1} \rangle
= \frac{1}{4 \pi^2} \int_0^{2 \pi} \int_0^{2 \pi} \int_{0}^{\pi} 
|(v_z)^{\gamma-1}| 
\sin \theta_b  d\theta_b d\phi_b d\phi_v = \frac{1}{\gamma}
\end{equation}

The coefficients for the excitation rates in the regime of 
$2<\gamma<3$ are more complicated as the dependence on 
$b/r_b$ cannot be factored out of the coefficient.  In addition to 
integrating over all angles, we must integrate over impact parameter.
For any $\gamma$, equation \ref{eqPintermediate} is written:

\begin{equation}
\label{eqPIntGeneral}
p(e) = \frac{G n_0 m_0 T_{\rm orb}}{e^{\gamma+1}}
\left( \frac{m_0}{m_b} \frac{v_b}{v_0} \right)^{\gamma-2}
\frac{\gamma-1}{2 \pi} V_{2-\gamma} \langle A_e^{\gamma-1} \rangle,
\end{equation}

\noindent
where $V_{\gamma-2}$ is discussed in section \ref{secColRate}; for a
Gaussian distribution of perturber velocities, $V_{\gamma-2}=2
\Gamma((1+\gamma)/2)$.  The term $\langle A_e^{\gamma-1} \rangle$
again contains the angular information.  Excitations for $2 < \gamma <
3$ are most important at $b \sim r_b$ so we can not assume that $\vec
b_2 \approx \vec b$. We introduce explicit notation for the the
components of the unit vector $\hat b_2 = b_{2,r} \hat r_b + b_{2,v}
\hat v_b + b_{2,n} \hat n$.  Then the angular average coefficient is:

\begin{equation}
\label{eqAvgAe}
\langle A_e^{\gamma-1} \rangle =
\frac{1}{8 \pi^2} 
\int_0^{2 \pi} \int_0^{2 \pi} \int_{0}^{\pi} \int_{0}^{\infty}
\left[ 
16 \left(\frac{b_{2,v}}{x_2}-\frac{b_v}{x_1} \right)^2 
+ 4 \left(\frac{b_{2,r}}{x_2}-\frac{b_{r}}{x_1}\right)^2
\right]^{(\gamma-1)/2} x_1 \sin \theta_b d x_1 d \theta_b d \phi_b d \phi_v,
\end{equation}

\noindent
with $x_1=b/r_b$ and $x_2=b_2/r_b$.  The magnitude and components 
of $\vec b_2$ are related to $\vec b$ and $\vec v_p$ as described 
in section \ref{secSingleEnc}: 
$\vec b_2 = \vec b - \vec r_b + \hat v_p (\vec r_b \cdot \hat v_p)$.
For $\gamma=25/12$ as discussed in \ref{secInt}, 
$\langle A_e^{13/12} \rangle \approx 15$.  For other $\gamma$ between 
2 and 3, this factor is of the same order, 10-15.

\bibliographystyle{apj}
\bibliography{ms}

\end{document}